\documentclass[letterpaper, 10 pt, conference]{ieeeconf}  

\IEEEoverridecommandlockouts                              

\usepackage{tikz}
\usepackage{amsmath, amsfonts, amssymb}
\usepackage{algorithmic}
\usepackage{tikz}
\let\labelindent\relax
\usepackage{enumitem}
\usepackage{multirow}
\usepackage{booktabs}
\usepackage{filecontents}
\usepackage{url}

\DeclareMathOperator*{\argmax}{arg\,max}

\title{\LARGE \bf
MADE: Malicious Agent Detection for Robust Multi-Agent Collaborative Perception
}

\author{Yangheng Zhao$^{*\dagger}$, Zhen Xiang$^{*\ddagger}$, Sheng Yin$^{*\dagger}$, Xianghe Pang$^{\dagger}$, Yanfeng Wang$^{\dagger}$, Siheng Chen$^{\mathparagraph\dagger}$
\thanks{$^{\dagger}$ Shanghai Jiao Tong University}%
\thanks{$^{\ddagger}$ University of Illinois, Urbana-Champaign}%
\thanks{$^{*}$ Equal contribution}
\thanks{$^{\mathparagraph}$ Corresponding to sihengc@sjtu.edu.cn }
}

\begin{document}

\maketitle
\thispagestyle{empty}
\pagestyle{empty}

\begin{abstract}

Recently, multi-agent collaborative (MAC) perception has been proposed and outperformed the traditional single-agent perception in many applications, such as autonomous driving.
However, MAC perception is more vulnerable to adversarial attacks than single-agent perception due to the information exchange.
The attacker can easily degrade the performance of a victim agent by sending harmful information from a malicious agent nearby.
In this paper, we propose \underline{M}alicious \underline{A}gent \underline{De}tection (MADE), a reactive defense specific to MAC perception that can be deployed by an agent to accurately detect and then remove any potential malicious agent in its local collaboration network.
In particular, MADE inspects each agent in the network independently using a semi-supervised anomaly detector based on a double-hypothesis test with the Benjamini-Hochberg procedure for false positive control.
For the two hypothesis tests, we propose a \textit{match loss} statistic and a \textit{collaborative reconstruction loss} statistic, respectively, both based on the \textit{consistency} between the agent to be inspected and the ego agent deployed with our detector.
We comprehensively evaluate MADE on a benchmark 3D dataset, V2X-sim, and a real-road dataset, DAIR-V2X, comparing it to baseline defenses.
Notably, with the protection of MADE, the drops in the average precision compared with the best-case `Oracle' defender are merely 1.27\% and 0.28\%, respectively.

\end{abstract}

\section{INTRODUCTION}\label{sec:introduction}

Through decades of efforts in computer vision and machine learning, single-agent perception has achieved remarkable success in object detection, tracking, and segmentation~\cite{centerpoint, pointpillars, ddetr, tracking, spvnas}.
However, the traditional single-agent perception still suffers from a number of inevitable limitations due to an individual perspective, such as occlusion and long-range issues~\cite{occluded, longrange}.
To this end, multi-agent collaborative (MAC) perception has been proposed recently, which allows multiple agents to share complementary perceptual information with each other, promoting a more holistic perception~\cite{liu2020when2com, wang2020v2vnet,li2021learning,where2comm}.

While existing studies on MAC perception mainly focus on improving perception performance~\cite{mac_review}, the associated security protocols are severely underexplored.
In particular, the information sharing between agents potentially creates more opportunities for adversaries to break the entire system.
For example, adversarial attacks (\cite{Szegedy_seminal}) have been extended to MAC object detection recently, where the performance of an arbitrary victim agent can be easily degraded by the adversary using an optimized feature map sent from a compromised agent in the collaboration network~\cite{tu_iccv_2021}.
Since generic defenses such as adversarial training cannot effectively mitigate this threat, the deployment of MAC perception tools remains risky in safety-critical scenarios due to potential economic losses and human fatalities~\cite{av_law, ai_bill}.

In this paper, we propose \underline{M}alicious \underline{A}gent \underline{De}tection (MADE), a reactive defense specific to adversarial attacks against MAC object detectors.
MADE leverages the nature of agent collaboration to accurately detect malicious agents in an ego agent's collaboration network and then remove them.
It is based on a multi-test using conformal p-values~\cite{pmlr-v25-vovk12, conformal_pv}, and with the Benjamini-Hochberg (BH) procedure for the false positive control~\cite{Benjamini1995}.
The multi-test is composed of two hypothesis tests on the \textit{consistency} between the ego agent and the agent to be inspected.
Specifically, we propose two novel detection statistics, a {\it match loss} statistic and a {\it collaborative reconstruction loss} statistic, as the consistent metrics for the two hypothesis tests, respectively.
An agent is claimed to be a malicious one if it is `inconsistent' with the ego agent as inferred by our multi-test.
Note that our MADE is different from generic proactive defenses since we not only protect agents from potential adversarial attacks but also help to catch the malicious entities launching the attack, which is socially impactful~\cite{cyber_law1, cyber_law2}.

Our main contributions are summarized as follows:

\begin{itemize}[leftmargin=*]

\item We propose MADE, a reactive, detection-based defense to address adversarial attacks against MAC object detectors by detecting and removing malicious agents in the local collaboration network.

\item We propose a multi-test detection framework for MADE, with two novel statistics, \textit{match loss} and \textit{collaborative reconstruction loss}, to assess the consistency between each putative malicious agent and the ego agent.

\item We comprehensively evaluate MADE on V2X-SIM and DAIR-V2X.
We show that MADE outperforms the previous state-of-the-art by recovering 22.13\% and 22.60\% average precision on the two datasets, respectively, merely 1.27\% and 0.28\% lower than the average precision achieved by the `Oracle' defender with ground-truth knowledge about the malicious agents, respectively.

\end{itemize}
\begin{figure*}
    \centering
    \includegraphics[width=17cm]{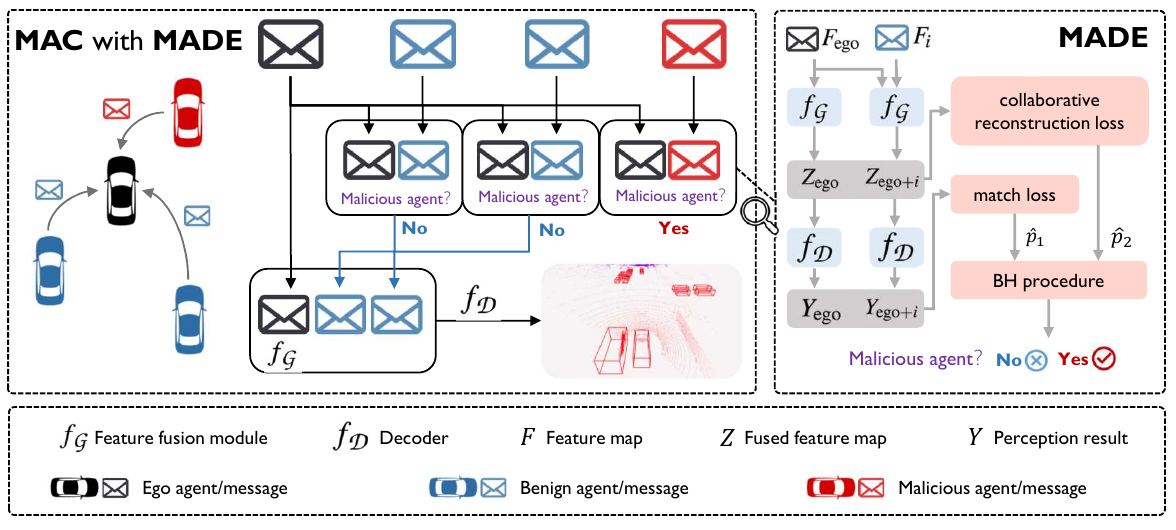}
    \caption{Illustration of a standard MAC perception pipeline protected by our proposed MADE.
    Message (i.e. intermediate feature map) from each agent will be inspected by MADE before fusing -- malicious agents will be removed once detected by MADE.
    Specifically, for any agent $a_i$ (with a feature map $F_i$) to be inspected, we obtain $Z_{\rm ego}$ directly from the ego agent's feature map $F_{\rm ego}$ (without fusion) and obtain a fused feature map $Z_{{\rm ego}+i}$ by fusing $F_i$ with $F_{\rm ego}$.
    We then obtain the bounding box proposals $Y_{\rm ego}$ and $Y_{{\rm ego}+i}$ from $Z_{\rm ego}$ and $Z_{{\rm ego}+i}$, respectively. A \textit{collaborative reconstruction loss} is computed on $Z_{\rm ego}$ and $Z_{{\rm ego}+i}$, while a \textit{match loss} statistic is computed on $Y_{\rm ego}$ and $Y_{{\rm ego}+i}$. Finally, conformal p-values are computed for the two tests respectively, which are then used for a multi-test with the BH procedure, leading to the inference result for agent $a_i$.}
    \label{fig:key_ideas_illustration}
\end{figure*}

\section{Related Work}\label{sec:related_work}

\subsection{MAC Perception}\label{subsec:background_multi_agent}

A MAC perception framework involves a group of agents, each deployed with the same model to produce its own perception output via collaboration with the other agents.
Different from other multi-agent learning paradigms, such as distributed learning focusing on training-time collaboration~\cite{chen2021distributed}, federated learning focusing on training data privacy~\cite{fl_seminal}, or multi-modal learning focusing on the aggregation of different sensory inputs \cite{multimodel}, MAC perception addresses the collaboration between agents \textit{during inference} \cite{bakliwal2018multi}.
In this paper, we consider state-of-the-art MAC object detectors that conduct the collaboration in the intermediate feature space, which achieves the best trade-off between communication cost and perception performance~\cite{where2comm, wang2020v2vnet, li2021learning, where2comm, xu2022v2x}.

\subsection{Adversarial Attacks and Defenses for MAC Perception}\label{subsec:background_mac_adv}

Adversarial (evasion) attacks originated from the study of the worst-case robustness of deep neural network (DNN) classifiers \cite{Szegedy_seminal}, and have been extensively studied for single-agent object detection~\cite{xie2017adversarial, adv_tshirt, hu2021natadv, adv_obj, Thys_2019_CVPR_Workshops, Li2018ExploringTV, dpatch, seeing}.
Recently, adversarial attacks have been extended to MAC object detection~\cite{tu_iccv_2021}.
The adversary applies an optimized perturbation to the feature map sent from a malicious agent to the victim agent to degrade its performance in bounding box prediction.
Such an attack is shown evasive to generic defenses against adversarial attacks, particularly the adversarial training approach~\cite{PGD}.

Prior to our work, a defense based on geometric consistency has been proposed to detect adversarial attacks against MAC object detectors~\cite{zhang2024cvp}.
However, this approach requires each agent to generate and share additional information only for the defense purpose (i.e., information not useful to MAC object detection); thus, it will not be further considered here since we do not allow the defender to manipulate the communication approach hardwired to the MAC object detector.
Closest to our work is the detection-based defense in~\cite{li2023us}, where potential malicious agents are identified using a consensus loss.
As will be shown experimentally, this defense is outperformed by our MADE, which leverages the power of two hypothesis tests with two different detection statistics, respectively.

\section{Threat Model}\label{sec:threat_model}

\subsection{MAC Object Detection Pipeline}\label{subsec:threat_model_pipeline}

We consider a collaboration network of $N$ agents $a_1, \cdots, a_N$.
For each agent, the input $X\in{\mathcal X}$ is the local sensory data and the output $Y=\{y^{(1)}, \cdots, y^{(L)}\}\in{\mathcal Y}$ is an ordered set of $L$ bounding box proposals.
Each bounding box proposal is a $k+d$ dimensional vector represented by $y=[p_1, \cdots, p_k, z_1, \cdots, z_d]^{\top}$, where $p_1, \cdots, p_k$ are the posteriors for all $k>1$ classes in the domain, respectively, and $z_1, \cdots, z_d$ jointly specify the spatial location of the bounding box.
We consider the state-of-the-art MAC object detector pipeline with the following three stages:
1) \textbf{Encoding stage.} For each agent $a_i$, an encoder $f_{\mathcal E}:{\mathcal X}\rightarrow{\mathcal F}$ is deployed to encode the local input $X_i$ captured by $a_i$ into a feature map $F_i\in{\mathcal F}$.
2) \textbf{Collaboration stage.} Through a communication channel, each agent $a_i$ receives the feature maps $\{F_j\}_{j\neq i}$ generated by all the other $(N-1)$ agents (using the same $f_{\mathcal E}$).
These received feature maps are aggregated with the local feature map $F_i$ using an asymmetrical function $f_{\mathcal G}:{\mathcal F}^N\rightarrow{\mathcal Z}$, which produces a fused feature map $Z_i=f_{\mathcal G}(F_i;\{F_j\}_{j\neq i})\in{\mathcal Z}$.
3) \textbf{Decoding stage.} The output $Y_i\in{\mathcal Y}$ of each agent $a_i$ is obtained by applying a decoder $f_{\mathcal D}:{\mathcal Z}\rightarrow{\mathcal Y}$ to its fused feature map $Z_i$.

\subsection{Adversarial Attack Against MAC Object Detector}\label{subsec:threat_model_attack}

We consider the adversarial attack in~\cite{tu_iccv_2021}, where $a_v$ denotes the victim agent targeted by the attacker, and $a_m$ ($m\neq v$) is the malicious agent controlled by the attacker.
For each agent $a_i$, the feature map obtained by applying the encoder $f_{\mathcal E}$ to its local input is denoted by $F_i\in{\mathcal F}$.
The attacker aims to degrade the performance of the victim agent $a_v$ by applying an optimized perturbation to the feature map $F_m$ sent from agent $a_m$ to $a_v$.
Let $Y'(\delta) = \{y'^{(1)}(\delta), \cdots, y'^{(L)}(\delta)\} = f_{\mathcal D} \circ f_{\mathcal G}(F_v; (F_m+\delta)\cup\{F_i\}_{i \neq v, i \neq m})$ be the $L$ bounding box proposals when $F_m$ is perturbed by $\delta$, the attacker aims to optimize $\delta$ by solving:
\begin{equation} \label{eq:attack_loss_aggregation}
\underset{||\delta||_p \leq \epsilon}{minimize} \quad \sum_{l=1}^L {\mathcal L}_{\text{adv}} (y^{(l)}, y'^{(l)}(\delta), \varrho(y^{(l)}))
\end{equation}
where $||\cdot||_p$ represents $\ell_p$ norm.
The categorization function $\varrho(\cdot):{\mathbb R}^{k+d}\rightarrow\{0, 1\}$ decides whether a bounding box proposal is associated with a `background' box (`0') or a `foreground' box associated with an object (`1').
The `per-box loss' in the sum of \eqref{eq:attack_loss_aggregation} is defined by:
\begin{equation*}
{\mathcal L}_{\text{adv}} (y, y', \varrho(y))= \begin{cases}
0 & \varrho(y)=0\\
-\log (1-p'_u) \cdot \operatorname{IoU}(y, y') & \varrho(y)=1
\end{cases}
\end{equation*}
where $u=\argmax_{c\in{\mathcal C}\setminus k}p_c$ is the largest confidence class (except the `background' class $k$) associated with the proposal $y$.
$p'_c$ is the posterior associated with proposal $y'$ for any class $c\in{\mathcal C}$. 
Following~\cite{tu_iccv_2021}, the attack optimization problem is solved by projected gradient descent (PGD)~\cite{PGD}.

While the procedure above corresponds to a single attacker with one malicious agent, in our experiments, we will consider both cases where there are multiple independent attackers and cases where one attacker controls multiple agents to launch a `collaborative' attack.

\section{Malicious Agent Detection Method}\label{sec:method}

\subsection{Overview}\label{subsec:key_ideas}

Without loss of generality, we nominally set the $N$-th agent $a_N$ to be a benign agent (dubbed the `ego' agent) deployed with our detector.
The goal here is to infer if any of the agents $a_1, \cdots, a_{N-1}$ is malicious, and then remove the detected agents from the local collaboration network.
However, there is no prior knowledge about the presence of any malicious agents or their identities.
The ground truth for the objects in the scene is also unavailable to the ego agent.
Moreover, no additional communication with other agents is allowed, except for the exchange of the feature maps, which is the default for MAC object detection.
The ego agent is only assumed with a set ${\mathcal D}_{\rm B}$ of (benign) feature maps (e.g.) collected locally from past instances.
These constraints make the detection problem very challenging.

Our key idea is to inspect each agent in $a_1, \cdots, a_{N-1}$ by assessing its `consistency' with the ego agent $a_N$ in both outputs and intermediate features.
Specifically, we propose a {\it match loss} statistic for the output-level consistency (see Sec. \ref{subsec:match_loss}) and a \textit{collaborative reconstruction loss} statistic for the intermediate-level consistency (see Sec. \ref{subsec:reconstruction_loss}).
While the two statistics can be leveraged for malicious agent detection independently through hypothesis testing, in this paper, we construct a multi-test to jointly exploit the power of the two statistics, with the false detection rate controlled by the Benjamini-Hochberg (BH) procedure~\cite{Benjamini1995} (see Sec. \ref{subsec:multi_test}).
The procedure of MADE is summarized in Fig. \ref{fig:key_ideas_illustration}.

\subsection{Match Loss}\label{subsec:match_loss}

The match loss statistic is an asymmetric metric designed to assess the consistency between two sets of bounding box proposals, say, $Y=\{y^{(1)}, \cdots, y^{(L)}\}$ and $Y'=\{y'^{(1)}, \cdots, y'^{(L)}\}$.
Let ${\mathcal I}_c$ and ${\mathcal I}'_c$ be the two sets of indices of all the proposed bounding boxes in $Y$ and $Y'$ that are predicted to some class $c\in{\mathcal C}$, respectively.
Let $\mathfrak{S}_c$ be all {\it matches} between ${\mathcal I}_c$ and ${\mathcal I}'_c$ that maps {\it each} $l\in{\mathcal I}_c$ to a {\it unique} $l'\in{\mathcal I}'_c$.
Note that to guarantee such uniqueness when $|{\mathcal I}'_c|<|{\mathcal I}_c|$, we pad ${\mathcal I}'_c$ to the same size as ${\mathcal I}_c$ with arbitrary `dummy' indices not appeared in $\{1, \cdots, L\}$ and associate each `dummy' index with an empty box $\varnothing$. Then, the {\it match loss} between $Y$ and $Y'$ is defined by:
\begin{equation*}\label{eq:match_loss_aggregation}
    {\mathcal L}_{\rm m}(Y, Y') = \sum_{c \in {\mathcal C}} \min_{\sigma \in \mathfrak{S}_{c}} \frac{1}{|{\mathcal I}_c|} \sum_{l\in {\mathcal I}_{c}} {\mathcal L}_{\rm box}(y^{(l)}, y'^{(\sigma(l))}; c)
\end{equation*}
where the optimal match for each class $c$ is solved using the Hungarian match algorithm~\cite{kuhn1955hungarian}.
Here, for any class $c\in{\mathcal C}$ and any two arbitrary bounding box proposals $y$ and $y'$,
\begin{equation}\label{eq:match_loss}
    {\mathcal L}_{\rm box}(y, y'; c) = \max (p_c - p'_c, 0) + \phi(1 - \operatorname{IoU}(z, z'))
\end{equation}
where $p_c$ and $p'_c$ represent the posteriors of class $c$ associated with bounding boxes $y$ and $y'$, respectively, and $\phi$ is the parameter balancing the two terms.

Focusing on the first term in Eq. (\ref{eq:match_loss}), a {\it large} loss will be induced only if $p'_c$ is much smaller than $p_c$, which is the usual case when the ego agent collaborates with a malicious agent.
By contrast, collaboration with a benign agent will not create a big decrement in $p'_c$.
Similarly, for the second term in Eq. (\ref{eq:match_loss}), collaboration with a malicious (or benign) agent will likely cause a large (or small) shift in the position of the proposed bounding boxes.
Thus, the match loss tends to be \textit{large} if the agent to be inspected is \textit{malicious}.

\subsection{Collaborative Reconstruction Loss}\label{subsec:reconstruction_loss}

We propose the collaborative reconstruction loss statistic to measure the consistency of the \textit{fused} feature maps of the ego agent with and without the collaboration with the agent $a_i$ to be inspected.
For convenience, these two fused feature maps are denoted as $Z_{{\rm ego}+i}=f_{\mathcal G}(F_{\rm ego};\{F_i, 0, \cdots, 0\})$ and $Z_{\rm ego}=f_{\mathcal G}(F_{\rm ego};\{0, \cdots, 0\})$ respectively.
If $a_i$ is a benign agent `consistent' with the ego agent, the \textit{residual feature map} $R_{{\rm ego}+i}=Z_{{\rm ego}+i}-Z_{\rm ego}$ will likely be less noisy than for $a_i$ being a malicious agent.
In other words, the residual feature maps associated with malicious agents and benign agents follow different distributions.

However, for our practical detection problem, only the feature maps associated with benign agents are available to the defender.
Thus, we propose to train an autoencoder on the residual feature maps for benign agents (simulated on ${\mathcal D}_{\rm B}$), such that residual feature maps for malicious agents will exhibit abnormally large reconstruction loss when going through the autoencoder.
Specifically, we denote the set of all ordered pairs of benign feature maps in ${\mathcal D}_{\rm B}$ as:
\begin{equation*}
{\mathcal F}_{\rm pair} ({\mathcal D}_{\rm B})=\{(F, F') | F\in{\mathcal D}_{\rm B}, F'\in{\mathcal D}_{\rm B}, F\neq F'\}
\end{equation*}
The training set is obtained from ${\mathcal F}_{\rm pair} ({\mathcal D}_{\rm B})$, denoted by:
\begin{equation*}\label{eq:ae_training_set}
\begin{aligned}
{\mathcal R} = \{&f_{\mathcal G}(F;\{F', 0, \cdots, 0\}) - f_{\mathcal G}(F;\{0, \cdots, 0\}) \\
&| (F, F')\in {\mathcal F}_{\rm pair}({\mathcal D}_{\rm B})\} 
\end{aligned}
\end{equation*}
Then the autoencoder $f_{\mathcal{AE}}: {\mathcal Z}\rightarrow{\mathcal Z}$ is trained by minimizing the average reconstruction loss on ${\mathcal R}$, i.e.:
\begin{equation*} \label{eq:ae_loss}
	\underset{\theta}{\text{minimize}} \quad \frac{1}{|{\mathcal R}|} \sum_{R\in{\mathcal R}} {\mathcal L}_{\rm{r}} (R, f_{\mathcal{AE}}(R; \theta))
\end{equation*}
where $\theta$ represents the parameters of the autoencoder $f_{\mathcal{AE}}$ and will be dropped in the sequel for brevity.
Our \textit{collaborative reconstruction loss} statistic for any residual feature map $R\in{\mathcal Z}$ is defined by the following mean square error:
\begin{equation*} \label{eq:recon_loss}
    {\mathcal L}_{\rm{r}} (R, f_{\mathcal{AE}}(R)) = ||R - f_{\mathcal{AE}}(R)||_2,
\end{equation*}
which tends to be large for malicious agents.

\subsection{Detection Inference Based on Multi-Test}\label{subsec:multi_test}

Our detection inference is based on {\it multi-test} using {\it conformal p-values} computed on calibration sets.
For each agent $a_i$ to be inspected, we first obtain the match loss statistic ${\mathcal L}_{\rm m}(Y_{\rm ego}, Y_{{\rm ego}+i})$ and the collaborative reconstruction loss statistic ${\mathcal L}_{\rm{r}} (R_{{\rm ego}+i}, f_{\mathcal{AE}}(R_{{\rm ego}+i}))$, respectively.
Then, our detection inference can be cast into the following multi-test:
\begin{equation}\label{eq:multi_test}
\begin{aligned}
    H_{1, 0}: (Y_{\rm ego}, Y_{{\rm ego}+i}) \sim P_{Y,Y'}&; \,\, H_{1, 1}: (Y_{\rm ego}, Y_{{\rm ego}+i}) \not\sim P_{Y,Y'}\\
    H_{2, 0}: R_{{\rm ego}+i} \sim P_R&; \,\, H_{2, 1}: R_{{\rm ego}+i} \not\sim P_R
\end{aligned}
\end{equation}
Here, $P_{Y,Y'}$ is the joint distribution for (output) random variables $Y=(f_{\mathcal D} \circ f_{\mathcal G})(F;\{0, \cdots, 0\})$ and $Y'=(f_{\mathcal D} \circ f_{\mathcal G})(F;\{F', 0, \cdots, 0\})$, and $P_R$ is the distribution for (residual feature map) random variable $R=f_{\mathcal G}(F;\{F', 0, \cdots, 0\}) - f_{\mathcal G}(F;\{0, \cdots, 0\})$, where $F$ and $F'$ are i.i.d. random variables for benign feature maps.

Since the ego agent is always benign, ideally, the rejection of any of these two hypotheses will be caused by $a_i$ being malicious.
Here, we employ the Benjamini-Hochberg (BH) procedure to control the overall false detection rate of the multi-test~\cite{fwer, fdr, Benjamini1995}.
We first obtain a p-value for each statistic under its associated null hypothesis independently.
Since both $P_{Y,Y'}$ and $P_R$ in (\ref{eq:multi_test}) are analytically intractable, for each statistic, we compute a conformal p-value on a {\it calibration set} obtained from ${\mathcal D}_{\rm B}$ by:
\begin{align*}
&\hat{p}_1 = \frac{1 + |\{ (Y, Y')\in{\mathcal J}_1^{{\rm cal}} | {\mathcal L}_{\rm m}(Y, Y') \geq {\mathcal L}_{\rm m}(Y_{\rm ego}, Y_{{\rm ego}+i}) \}|} {1 + |{\mathcal J}_1^{{\rm cal}}|}
\\
&\hat{p}_2 = \\
&\frac{1 + |\{ R\in{\mathcal J}_2^{{\rm cal}} | {\mathcal L}_{\rm{r}} (R, f_{\mathcal{AE}}(R)) \geq {\mathcal L}_{\rm{r}} (R_{{\rm ego}+i}, f_{\mathcal{AE}}(R_{{\rm ego}+i})) \}|} {1 + |{\mathcal J}_2^{{\rm cal}}|}
\end{align*}
where subscripts `1' and `2' correspond to the match loss and the collaborative reconstruction loss respectively and ${\mathcal J}_i^{{\rm cal}}$ denotes the calibration set.
Then, following the BH procedure, $a_i$ is deemed a malicious agent if:
\begin{equation*}\label{eq:bh_inference}
\min \{\hat{p}_1, \hat{p}_2\} \leq \frac{\alpha}{2} \quad {\rm or} \quad \max \{\hat{p}_1, \hat{p}_2\} \leq \alpha
\end{equation*}
where $\alpha$ is the desired FPR (e.g., the classical $\alpha=0.05$ for statistical testing).

\section{Experiments}\label{sub:exp}

\begin{figure*}[h!]
	\centering
	\includegraphics[width=0.8\linewidth]{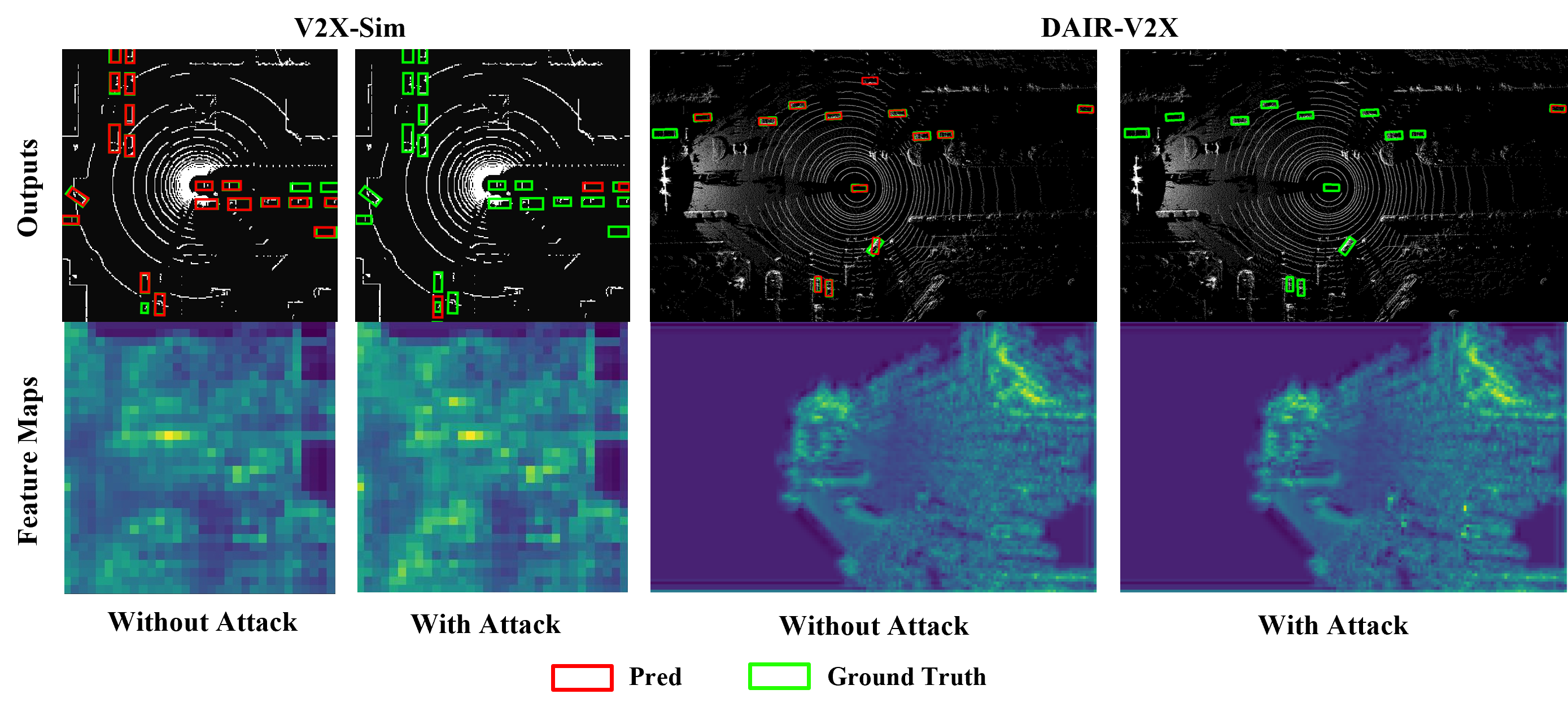}
	\caption{Comparative visualization of intermediate feature maps and detection outputs with and without adversarial attack}
	\label{fig:vis}
\end{figure*}

\begin{table*}[t!]
    \caption{End-to-end defense performance of MADE compared with the baseline.}
    \centering{
        \scalebox{1.0}{%
            \begin{tabular}{cc|cccc|cccc}
                \toprule \hline
                \multirow{2}*{} & \multirow{2}*{} & \multicolumn{4}{c}{V2X-SIM} & \multicolumn{4}{c}{DAIR-V2X}\\
                \cmidrule(lr){3-6} \cmidrule(lr){7-10}
                & & no attack & $\epsilon=0.05$ & $\epsilon=0.1$ & $\epsilon=0.2$ & no attack & $\epsilon=1.0$ & $\epsilon=1.5$ & $\epsilon=2.0$\\
                \hline
                \multirow{2}{*}{(reference)} & no defense & 61.82 & 53.30 & 36.36 & 22.40 & 64.97 & 47.40 & 36.55 & 27.99 \\
                & oracle & 61.82 & 59.76 & 59.76 & 59.76 & 64.97 & 59.43 & 59.43 & 59.43\\  \hline
                \multirow{1}{*}{(baseline)} 
                & ROBOSAC~\cite{li2023us} & 59.54 & 54.13 & 52.35 & 53.97 & 64.97 & 47.75 & 42.19 & 42.84\\
                \multirow{3}{*}{\textbf{(ours)}} & \textbf{ML only} & 61.91 & 56.56 & 57.72 & 57.79 & 64.87 & 57.11 & 58.48 & 58.95\\
                & \textbf{CRL only} & 60.38 & 54.77 & 54.47 & 50.35 & 60.16 & 57.99 & 57.47 & 58.04\\
                & \textbf{MADE (ML+CRL)} & 61.10 & \textbf{56.91} & \textbf{58.49} & \textbf{58.40} & 60.58 & \textbf{58.45} & \textbf{59.15} & \textbf{59.28}\\
                \hline
                \bottomrule
            \end{tabular}
        }
    }
\label{tab:defense_ap}
\end{table*}

\subsection{Experimental Setup}\label{subsec:exp_settings}

\textbf{Dataset.} We conduct experiments on two benchmarks, a vehicle-to-everything dataset V2X-SIM and a real-road vehicle-to-infrastructure dataset DAIR-V2X~\cite{li2021learning, yu2022dair}.
For V2X-SIM, we conduct an out-of-distribution evaluation by reserving all 8 scenes with 2 or 3 agents for the defender for detection, while 3 scenes with 4 agents are used for performance evaluation.
DAIR-V2X involves two agents (vehicle and infrastructure) in each scene.
We use 4800 frames to train the MAC object detector, with the remaining 1800 frames for our detection method and evaluation.\\
\textbf{MAC object detector.} For V2X-SIM, we use a pre-trained DiscoNet~\cite{li2021learning}, where the local input LiDAR point cloud for each agent is encoded to $32\times32$ bird eye view (BEV) feature maps with 256 channels before communication.
For DAIR-V2X, we train a multi-scale Where2comm with temperature 20 \cite{where2comm}, where the LiDAR point cloud for each agent is encoded to $100 \times 252$, $50 \times 126$, and $25 \times 63$ feature maps with 64, 128, and 256 channels, respectively.\\
\textbf{Attack details.} For V2X-SIM with four agents per frame, we generate four attack instances with the four agents taking turns as the ego agent; and the malicious agent is randomly chosen from the remaining three agents.
For DAIR-V2X with only two agents, we generate one attack instance for each frame by setting the vehicle as the ego agent and the infrastructure as the malicious agent.
In each attack instance, we manipulate the feature map sent from the malicious agent to the ego agent following our threat model in Sec. \ref{subsec:threat_model_attack}.
We consider maximum perturbation sizes (measured by the $\ell_{\infty}$ norm of the perturbation $\delta$) $\epsilon\in\{0.05, 0.1, 0.2\}$ and $\epsilon\in\{1.0, 1.5, 2.0\}$ for V2X-SIM and DAIR-V2X, respectively.
PGD is conducted for 10 steps with step size $\epsilon/5$ for V2X-SIM and 80 steps with step size $\epsilon/30$ for DAIR-V2X.
In Fig. XXX, we show the visualization of the feature maps being manipulated by the adversary for V2X-SIM and DAIR-V2X for attacks with $\epsilon=0.2$ and $\epsilon=2.0$ respectively.
Our created attacks for evaluation are stealthy, with almost imperceptible artifacts on the manipulated feature maps.\\
\textbf{Details of MADE.} In each attack instance, we inspect all the agents other than the ego agent.
The match loss is computed with the multiplier set to $\phi = 1.0$ in Eq. \ref{eq:match_loss} and the calibration sets containing 2800 and 1000 pairs of (benign) output bounding box proposals for V2X-SIM and DAIR-V2X, respectively.
For the collaborative reconstruction loss, the training sets of the autoencoders for V2X-SIM and DAIR-V2X contain 23700 and 3811 instances, respectively.
For V2X-SIM, we use U-Net \cite{unet} as the architecture of the autoencoder.
Training is performed using Adam optimizer for 200 epochs with batch size 256 and learning rate 0.0001 (with a 0.1 decay per 50 epochs).
For DAIR-V2X, we use the same architecture in~\cite{ae_architecture} for the autoencoder.
Training is performed using Adam optimizer for 200 epochs with batch size 64 and learning rate 0.001 (with a 0.1 decay per 50 iterations).
Finally, we set the desired FPR to the classical $\alpha=0.05$ for the multi-test inference procedure.\\
\textbf{Evaluation metrics.} We evaluate the effectiveness of defenses based on the end-to-end performance of the object detector measured by the average precision (AP) at 0.5 IoU (denoted by AP@0.5 and in percentage).

\subsection{Defense Performance Evaluation}\label{subsec:exp_defense}

In Tab. \ref{tab:defense_ap}, we show the end-to-end AP@0.5 of our MADE against the attacks we created for evaluation, compared with the previous state-of-the-art approach ROBOSAC~\cite{li2023us}.
We also consider two single-test variants of MADE with the match loss (ML) statistic and the collaborative reconstruction loss (CRL) statistic respectively.
Moreover, the worst-case performance when there is no defense and the best-case performance when the defender is an `Oracle' with ground truth knowledge that can accurately identify and remove all malicious agents are shown for reference.
As shown in the table, both defenses with a single test on our proposed ML or CRL statistic perform well with higher restored APs compared with the baseline for almost all attacks.
Notably, the defense that only uses our ML statistic has already outperformed the baseline with uniformly higher restored APs.
Furthermore, our MADE defense performs even better than the two single-test defenses against the two attacks on the two datasets by leveraging the detection power of both ML and CRL statistics, with the recovered AP@0.5 close to the best-case `Oracle' defender for both datasets.

\begin{table}[t!]
    \setlength{\tabcolsep}{3.7pt}
    \caption{Malicious agent detection accuracy of MADE on V2X-Sim.}
    \centering{
        {%
            \begin{tabular}{c|cccccc}
                \toprule \hline
                & \multicolumn{2}{c}{$\epsilon=0.05$} & \multicolumn{2}{c}{$\epsilon=0.1$} & \multicolumn{2}{c}{$\epsilon=0.2$} \\
                \cmidrule(lr){2-3} \cmidrule(lr){4-5} \cmidrule(lr){6-7}
                & TPR & FPR & TPR & FPR & TPR & FPR\\
                \hline
                ML only & 0.667 & 0.150 & 0.963 & 0.138 & 0.965 & 0.124 \\
                CRL only & 0.255 & 0.025 & 0.733 & 0.018 & 0.825 & 0.015 \\
                MADE (ML+CRL) & 0.593 & 0.076 & 0.958 & 0.069 & 0.947 & 0.055 \\
                \hline
                \bottomrule
            \end{tabular}
        }
    }
\label{tab:defense_tpr_fpr}
\end{table}

In addition, we demonstrate the true positive rate (TPR) and the false positive rate (FPR) of our MADE.
Specifically, TPR is defined as the proportion of malicious agents being successfully detected, while FPR is defined as the proportion of benign agents being incorrectly detected.
As shown in Tab \ref{tab:defense_tpr_fpr}, both ML and CRL can be used to effectively distinguish malicious agents from benign ones independently, while our MADE combining the two statistics is the only method achieving both high TPR and controlled FPR close to the prescribed significance level $\alpha=0.05$.

\subsection{Ablation Study}\label{subsec:ablation}

In this section, we conduct an ablation study on the parameter $\phi$ in match loss, which balances the classification difference (the first term in Eq.(\ref{eq:match_loss})) and the localization difference (the second term in Eq.(\ref{eq:match_loss})).
As shown in Tab. \ref{tab:ablation}, for both datasets, the performance of MADE is generally insensitive to the choice of $\phi$ as long as it is close to 1.0, which is our default setting.


\begin{table}[]
    \caption{Ablation study on parameter $\phi$ in match loss.}
    \setlength{\tabcolsep}{2.7pt}
    \centering{{%
        \begin{tabular}{cc|ccccc}
            \toprule \hline
            \multirow{2}*{} & \multirow{2}*{} & $\phi = 0.01$ & $\phi = 0.5$ & $\phi = 1$ & $\phi = 2$ & $\phi = 100$ \\
            \hline
            \multirow{2}{*}{V2X-SIM} & attack & 58.19 & 58.25 & 57.72 & 57.13 & 54.49 \\ & no attack & 61.73 & 60.75 & 61.91 & 60.10 & 60.01    \\
            \hline
            \multirow{2}{*}{DAIR-V2X} & attack & 59.41 & 59.12 & 58.48 & 57.84 & 54.47 \\ & no attack & 65.04 & 64.93 & 64.87 & 64.85 & 64.81    \\
             \hline
            \bottomrule
        \end{tabular}
        }
    }
\label{tab:ablation}
\end{table}

\subsection{Detection Performance against Adaptive Attacks}\label{subsec:exp_adaptive}

Here, we consider more challenging scenarios with multiple malicious agents.
In particular, there may be more than one attacker in the scene, and each attacker controls a malicious agent to launch an attack against the (ego) victim agent \textit{independently}.
Moreover, there may exist one attacker who controls multiple malicious agents to launch an attack \textit{collaboratively} by jointly optimizing the feature maps sent from these malicious agents to the victim agent.
Such a collaborative attack involving more than one agent can be regarded as an adaptive attack to our proposed defense.
This is because we inspect each agent other than the ego agent independently, while for multiple malicious agents in collaboration, the performance degradation at the ego agent induced by a single malicious agent may not be severe enough to be detected, e.g., using the match loss statistic.

\begin{table}[t!]
    \setlength{\tabcolsep}{5pt}
    \caption{Performance of ML, CRL, and MADE against attacks with multiple independent attackers and attacks with multiple collaborative malicious agents on V2X-SIM.}
    \vspace{0.075in}
    \centering{
        \scalebox{1.0}{%
            \begin{tabular}{c|cc|cc}
                \toprule \hline
                & \multicolumn{2}{c}{Independent} & \multicolumn{2}{c}{Collaborative}\\
                \cmidrule(lr){2-3} \cmidrule(lr){4-5}
                No. malicious & 2 agents & 3 agents & 2 agents & 3 agents\\ \hline
                no defense & 21.55 & 15.76 & 20.20 & 14.35  \\
                oracle & 57.40 & 49.56 & 57.40 & 49.56 \\ \hline
                ML only & 55.33 & 48.68 & 53.57 & 45.47 \\
                CRL only & 50.96 & 42.69 & 41.09 & 27.48\\
                MADE & 55.98 & 48.51 & 53.79 & 43.97  \\
                \hline
                \bottomrule
            \end{tabular}
        }
    }
\label{tab:defense_adaptive}
\end{table}

In Tab. \ref{tab:defense_adaptive}, we show the end-to-end object detection performance (measured by AP@0.5) of ML, CRL, and the proposed MADE, against two and three coexisting but independent attackers, respectively, and two and three collaborative malicious agents, respectively.
Here, we consider the V2X-SIM dataset only since the DAIR-V2X dataset has only two agents in each scene.
With more malicious agents (either independent or collaborative) and fewer benign agents, the object detection performance of the victim agent is clearly degraded.
Our MADE achieves 55.98\% and 48.51\% AP@0.5, merely 1.42\% and 1.05\% lower than the best-case ``Oracle'' defender, respectively.
For the collaborative attacks, there is a slight drop in the performance of our MADE, but it still achieves 53.79\% and 43.97\% AP@0.5, merely 3.61\% and 5.59\% lower than the best-case ``Oracle'' defender, respectively.

\subsection{Unsupervised Adaption}\label{subsec:exp_unsupervised}

Here, we adapt our proposed MADE defense to address a more challenging \textit{unsupervised} detection scenario where past information is not allowed.
The key intuition is that the manipulation of a feature map sent from a malicious agent is specific to the victim agent, which will not clearly affect the outputs of the other agents.
Thus, if the ego agent (with output $Y_{\rm ego}$) is indeed a victim agent targeted by one or more attackers, there will exist some agent $a_i$ with output $Y_i$, such that the match loss ${\mathcal L}_{\rm m}(Y_{\rm ego}, Y_i)$ is a large outlier to the match losses computed for non-ego agents.
For an instance with four agents $a_1, \cdots, a_4$ where $a_1$ is the ego agent, for example, we aim to assess the atypicality of ${\mathcal L}_{\rm m}(Y_1, Y_2)$, ${\mathcal L}_{\rm m}(Y_1, Y_3)$, and ${\mathcal L}_{\rm m}(Y_1, Y_4)$, with reference to the other nine match losses (`centered' at the three non-ego agents).
To achieve this in an unsupervised fashion, we apply a robust detection rule based on median absolute deviation (MAD) \cite{MAD}.
We first compute the MAD of the ML statistics for all ordered pairs of agents in the scene.
Then, we compute a score for each match loss $\mathcal{L}$ of the ego agent by:
\begin{equation*}
S = (\mathcal{L} - {\rm median}) / (1.4826\cdot{\rm mad})
\end{equation*}
where the median is taken over all the match loss statistics, and the constant 1.4826 is associated with a 0.95 significance under the Gaussian assumption.
The agent associated with the match loss is deemed to be a malicious agent if the score $S$ is greater than 2.
In Tab. \ref{tab:defense_unsupervised}, we show the performance of the unsupervised adaption of our method against attacks with one, two, and three malicious agents on V2X-SIM with and without collaboration.
For this challenging defense task, our unsupervised adaption achieves a decent performance for attacks with one malicious agent and some effect when there are more malicious agents.

\begin{table}[t!]
    \setlength{\tabcolsep}{3pt}
    \caption{Performance of the unsupervised adaption of MADE.}
    \centering{
        \scalebox{1.0}{%
            \begin{tabular}{cc|cc|cc}
                \toprule \hline
                & & \multicolumn{2}{c}{Independent} & \multicolumn{2}{c}{Collaborative}\\
                \cmidrule(lr){3-4} \cmidrule(lr){5-6}
                No. mal. & 1 agent & 2 agents & 3 agents & 2 agents & 3 agents\\ \hline
                AP@0.5 & 45.89 & 34.90 & 23.85 & 35.79 & 21.85 \\
                \hline
                \bottomrule
            \end{tabular}
        }
    }
\label{tab:defense_unsupervised}
\end{table}

\section{CONCLUSIONS}

This paper focuses on the security issues of the emergent MAC perception systems.
We propose a reactive defense that detects and removes malicious agents in the collaboration network of the ego agent.
Our defense is based on a multi-test that involves two novel statistics, match loss and collaborative reconstruction loss, with the Benjamini-Hochber for false detection rate control.
Comprehensive evaluations on both simulation and real-world benchmarks show that our defense is more effective than the previous state-of-the-art.
It achieves 58.49\% and 59.15\% AP@0.5 on V2X-SIM and DAIR-V2X, respectively, merely 1.27\% and 0.28\% lower than AP@0.5 achieved by the `Oracle' defender, respectively.









\bibliographystyle{plain}
\bibliography{ref}

\end{document}